\documentclass[12pt,aps,nofootinbib,preprintnumbers]{revtex4-2}
\usepackage{amssymb}
\usepackage{amsmath}
\usepackage[colorlinks=true, pdfstartview=FitV, linkcolor=blue, citecolor=blue, urlcolor=blue]{hyperref}

\newcommand{\bx}{{\boldsymbol x}}
\newcommand{\bp}{{\boldsymbol p}}

\newcommand{\bu}{{\boldsymbol u}}
\newcommand{\bnabla}{{\boldsymbol \nabla}}
\newcommand{\bomega}{{\boldsymbol \omega}}
\newcommand{\bzero}{{\boldsymbol 0}}
\newcommand{\bK}{{\boldsymbol K}}
\newcommand{\bZ}{{\boldsymbol Z}}
\newcommand{\bJ}{{\boldsymbol J}}

\newcommand{\hp}{{\hat p}}

\newcommand{\calC}{\mathcal{C}}
\newcommand{\calD}{\mathcal{D}}
\newcommand{\calG}{\mathcal{G}}
\newcommand{\calH}{\mathcal{H}}

\newcommand{\calK}{\mathcal{K}}
\newcommand{\calZ}{\mathcal{Z}}
\newcommand{\calA}{\mathcal{A}}
\newcommand{\calS}{\mathcal{S}}
\newcommand{\calT}{\mathcal{T}}

\newcommand{\one}{ {(1)} }
\newcommand{\dis}{\displaystyle}

\newcommand{\rme}{\mathrm{e}}
\newcommand{\rmd}{\mathrm{d}}
\newcommand{\rmi}{\mathrm{i}}

\newcommand{\rout}{\bgroup \color{red} \ULdepth=-.5ex \ULset}
\newcommand{\bout}{\bgroup \color{blue} \ULdepth=-.5ex \ULset}

\begin{document}

\title{Photonic spin Hall effect \\ from quantum kinetic theory in curved spacetimes}
\author{Kazuya~Mameda}
\affiliation{Theoretical Research Division, Nishina Center, RIKEN, Wako, Saitama 351-0198, Japan} 
\affiliation{Department of Physics, Tokyo University of Science, Tokyo 162-8601, Japan}
\author{Naoki~Yamamoto}
\affiliation{Department of Physics, Keio University, Yokohama 223-8522, Japan}
\author{Di-Lun~Yang}
\affiliation{Institute of Physics, Academia Sinica, Taipei, 11529, Taiwan}
\preprint{RIKEN-QHP-494}

\begin{abstract}
	Based on quantum field theory, we formulate the Wigner function and quantum kinetic theory for polarized photons in curved spacetimes which admit a covariantly constant timelike vector.
	From this framework, the photonic chiral/zilch vortical effects are reproduced in a rigidly rotating coordinate.
	In a spatially inhomogeneous coordinate, we derive the spin Hall effect for the photon helicity current and energy current in equilibrium.
	Our derivation reveals that such photonic Hall effect are related to the photonic vortical effect via the Lorentz invariance and
	their transport coefficients match each other.
\end{abstract} 

\maketitle

\section{Introduction}
The geometric phase or Berry phase~\cite{Berry1984} is one of the most important concepts of quantum transport in modern physics.
For electrons in solid state physics, Berry curvature generates various intriguing phenomena, such as the anomalous Hall effect, anomalous Nernst effect, and thermal Hall effect~\cite{Xiao:2009rm}. The so-called chiral magnetic effect~\cite{Vilenkin:1980fu,Nielsen:1983rb,Fukushima:2008xe} and chiral vortical effect~\cite{Vilenkin:1979ui,Son:2009tf,Landsteiner:2011cp} can also be described by the Berry curvature intrinsic to chirality/helicity of relativistic fermions~\cite{Son:2012wh,Stephanov:2012ki}. Due to the similar topological nature of photons, some transport phenomena are shared by both fermions and photons. For instance, rotating photons realize a counterpart of the fermionic chiral vortical effect~\cite{Avkhadiev:2017fxj,Yamamoto:2017uul,PhysRevA.96.043830,Huang:2018aly,Prokhorov:2020okl} and zilch vortical effect~\cite{Chernodub:2018era,Copetti:2018mxw}.
Other prominent photonic transport phenomena due to the Berry phase is the spin Hall effect, which is also referred to as optical Magnus effect~\cite{PhysRevA.46.5199}.
A source of this effect is refraction in a medium or a background gravitational field.
Along these lines, the photonic spin Hall effect has been studied based on the geometrical optics or the wave-packet theory~\cite{PhysRevLett.93.083901,Berard:2004xn,Gosselin:2006wp}, Fermat principle with Berry curvature~\cite{BLIOKH2004181,Duval:2005ky}, the path-integral formalism~\cite{Yamamoto:2017gla} and the WKB approximation~\cite{Oancea:2020khc,Andersson:2020gsj}.

Despite various studies, the photonic spin Hall effect has not been derived from the underlying quantum electrodynamics (QED).
One systematic way to derive a kinetic theory for photons based on QED is the Wigner function approach~\cite{Huang:2020kik,Hattori:2020gqh,Lin:2021mvw,Hidaka:2022dmn}.
In this approach, the kinetic theory is derived from the equations of motion for Green's functions, and we can respect the covariance of kinetic theory and take into account collisional effects; for the application to chiral fermions, see Ref.~\cite{Hidaka:2016yjf}.
The extension to curved spacetime enables us to field-theoretically derive the photonic spin Hall effect induced by a gravitational field that can be regarded as an effective refractive index of a medium.
The resulting transport theory is also applicable to various photonic systems under genuine gravitational fields or with a thermal gradient or spatial inhomogeneity, as in the fermionic case~\cite{Liu:2018xip,Liu:2020flb}.

In this paper, we formulate the photonic quantum kinetic theory in curved spacetimes which admit a covariantly constant timelike vector, using the Wigner function formalism. 
We apply our framework to two specific geometries.
First, we show that the chiral and zilch vortical effects are reproduced in a rigidly rotating coordinate.
This is a demonstration of the validity of our framework.
Second, the photonic spin Hall effect is reproduced in the coordinate with a spatial inhomogeneity.
At the same time, we derive the energy flow of polarized photons due to the refraction or in the background gravitational field, that is, the photonic spin Hall energy current. One of the important observations here is that the photonic spin Hall effect are related to the photonic vortical effect through the Lorentz invariance, and as a consequence, their transport coefficients match.

This paper is organized as follows: In Sec.~\ref{sec:Wigner}, we derive the Wigner function with quantum corrections and free-streaming kinetic theory for polarized photons. In Sec.~\ref{sec:currents}, we compute various photonic currents, such as the chiral/zilch vortical effect induced by the vorticity and the spin Hall effect in the curved geometry mimicking a medium with an inhomogeneous refractive index. 
In Sec.~\ref{sec:discussion}, we conclude our work with discussions.

In this paper, the symmetrized and antisymmetrized tensor products are written as $X^{(\mu}Y^{\nu)} = \frac{1}{2}(X^\mu Y^\nu + X^\nu Y^\mu)$ and $X^{[\mu}Y^{\nu]} = \frac{1}{2}(X^\mu Y^\nu - X^\nu Y^\mu)$, respectively.%
\footnote{This convention is different from that in Ref.~\cite{Hattori:2020gqh} by the factor of $1/2$.}
We will set $c = k_{\rm B} = 1$ and keep $\hbar$ only to indicate the $\hbar$ expansion, but we will suppress other $\hbar$'s.%
\footnote{
Let us comment on implicit $\hbar$ factors.
The first is the factor of $\hbar^{-3}$ from the momentum space measure $\int_\bp$.
The second is the one from the definition of the photon Wigner function $W_{\mu\nu}$ in Eq.~(\ref{eq:W}): one can check that in order to reproduce its correct physical dimension, $W_{\mu\nu}$ needs to be multiplied by the additional factor of $\hbar^{2}$. For all quantities calculated with the Wigner function in this paper, hence we implicitly multiply $\hbar^{-1}$ in total.%
}

\section{Wigner function}\label{sec:Wigner}
In this section, we formulate the photonic quantum transport theory in curved spacetime.
The basic strategy is the same as Ref.~\cite{Liu:2018xip}.
From the viewpoint of quantum field theory, the transport theory is described by the Wigner function.
For the U(1) gauge field $A_\mu(x)$ in curved spacetime, the Wigner transformation of the lesser propagator is defined as
\begin{equation}
 \label{eq:W}
 \begin{split}
  W_{\mu\nu}^h (x,p) = \int_y \rme^{-\rmi p\cdot y/\hbar} \langle A^h_\nu(x,\tfrac{y}{2}) A^h_\mu(x,-\tfrac{y}{2}) \rangle \,,
 \end{split}
\end{equation}
where the momentum variable $p_\mu$ (its conjugate variable $y^\mu$) is a covariant (contravariant) vector, and we define $\int_y = \int \rmd^4y\,\sqrt{-g}$ and $g=\det(g_{\mu\nu})$.
The superscript $h=\mathrm{R}, \mathrm{L}$ corresponds to the right- and left-handed helicity.
In the above equation, $A_\mu(x,y)$ is a covariantly translated field operator:
\begin{equation}
\label{eq:trans}
 A_\mu (x,y)
 = \biggl( 1 + y^\alpha \nabla_\alpha + \frac{1}{2} y^\alpha y^\beta \nabla_\alpha \nabla_\beta + \cdots \biggr) A_\mu (x)\,,
\end{equation}
where $\nabla_\mu A_\nu = \partial_\mu A_\nu - \Gamma^\rho_{\mu\nu} A_\rho$ with $\Gamma^\rho_{\mu\nu} = \frac{1}{2} g^{\rho\alpha}(\partial_\mu g_{\nu\alpha} + \partial_\nu g_{\mu\alpha}-\partial_\alpha g_{\mu\nu})$.
This keeps the Wigner function covariant under general coordinate transformation.
For latter convenience, let us introduce the following derivative operator:
\begin{equation}
\label{eq:lift_y}
 D_\mu \Phi(x,y)
   = (\nabla_\mu - \Gamma^\lambda_{\mu\nu} y^\nu\partial^y_\lambda) \Phi(x,y) \,,
\end{equation}
where $\partial_\lambda^y = \frac{\partial}{\partial y^\lambda}$ is the vertical derivative on the tangent bundle and $\Phi(x,y)$ is an arbitrary tensor field.
Thanks to the useful property $[D_\mu,y_\mu]=0$, the translation~\eqref{eq:trans} can be written as%
\footnote{Strictly speaking, this is the exponential map on the tangent bundle, and is valid only in geodesically complete spacetimes~\cite{derezinski2020pseudodifferential}.}
\begin{equation}
  A_\mu (x,y) = \exp(y\cdot D)A_\mu(x) \,.
\end{equation}
We also introduce a similar derivative of functions on the momentum space as
\begin{equation}
\label{eq:lift_p}
  D_\mu \Psi(x,p)
   = (\nabla_\mu + \Gamma^\lambda_{\mu\nu} p_\lambda\partial_p^\nu ) \Psi(x,p)\,,
\end{equation}
where $\partial^\nu_p = \frac{\partial}{\partial p_\nu}$ is the vertical derivative on the cotangent bundle and $\Psi(x,p)$ is an arbitrary tensor field.
In momentum space, we have $[D_\mu, p_\nu] = 0$.
In geometry, the derivative $D_\mu$ in Eqs.~\eqref{eq:lift_y} and~\eqref{eq:lift_p} is called the horizontal lift of $\nabla_\mu$.

Let us write down the transport equation of the free streaming photons.
We hereafter ignore the Riemann curvature contributions.
This simplification is always justified when we focus on the $O(\hbar)$ perturbation theory. In this paper, we will work out the kinetic theory, Wigner function, and associated effects up to $O(\hbar)$ as the leading-order quantum corrections in the $\hbar$ expansion.%
\footnote{The Wigner transformation in Eq.~(\ref{eq:W}) may lead to an overall factor of $\hbar$ that appears in the denominator, while this does not affect the power counting and one should only focus on the $\hbar$ expansion for the numerator.}
Through almost the same steps as the fermionic case~\cite{Liu:2018xip}, the equation of motion of the photonic Wigner function is found to be
\begin{equation}
 \label{eq:Weq}
    \bigl(
  		g^{\mu\nu}g^{\lambda\rho} 
  		- g^{\lambda\mu}g^{\nu\rho}
  	\bigr)
   		\biggl(p_{\mu} + \frac{\rmi \hbar}{2} D_{\mu}\biggr)
  		\biggl(p_{\nu} + \frac{\rmi \hbar}{2} D_{\nu}\biggr)
  		W^h_{\lambda\tau} 
  = O(\hbar^2) \,.
\end{equation}
The derivation of the above equation is shown in the Appendix.
It is readily checked that the flat spacetime limit of the above equation reproduces that derived in Ref.~\cite{Hattori:2020gqh}.

It is notable to make a comparison with Ref.~\cite{Bildhauer:1989dp}, where the photonic kinetic theory in curved spacetime is obtained based on the symmetric properties of the Wigner function.
We have confirmed that Eq.~\eqref{eq:Weq} is reproduced from the zero Riemann curvature limit of Ref.~\cite{Bildhauer:1989dp} [see Eq.~(4.1) therein].
The main goal of Ref.~\cite{Bildhauer:1989dp} is to derive the transport equation of the Wigner function, which however does not solve for a perturbative solution of the Wigner function with the $\hbar$ expansion. 
In the later part of this paper, we further derive such a solution for the Wigner function giving rise to the spin Hall effect under a particular type of spacetime geometry.

In order to further eliminate redundant degrees of freedom for Eq.~\eqref{eq:Weq}, we employ a typical gauge choice.
In the flat spacetime, one of the advantageous choices is the Coulomb gauge $\partial_\perp^\mu A^h_\mu = 0$.%
\footnote{Another is the Lorentz gauge $\partial^\mu A^h_\mu = 0$, with which the photonic quantum transport is discussed in Ref.~\cite{Huang:2020kik}.}
Here $\perp$ denotes the transverse projection to a vector $n^\mu$, which specifies the Lorentz frame for spin polarization of photons.
Besides, when we adopt the rest frame vector $n_\mu = (1,\bzero)$, the whole computation becomes simpler thanks to the condition $\partial_\mu n_\nu = 0$~\cite{Hattori:2020gqh}.
We should emphasize that physical quantities are independent of both the gauge and frame choices.
In curved spacetime, the corresponding frame vector is normal to the spatial hypersurface where time is constant. That is,
\begin{equation}
\label{eq:frame}
 n_\mu = ((g^{00})^{-1/2},\bzero) \,,
\end{equation}
which is normalized as $n^2 =1$.
Hence, we may extend the same strategy as in flat spacetime if a a given metric tensor $g_{\mu\nu}$ satisfies
\begin{equation}
\label{eq:nabla_n}
  \nabla_\mu n_\nu = 0 \,.
\end{equation}
This equation means that the spacetime admits covariantly constant timelike vector fields.
For instance, the Einstein static universe belongs to the class of such spacetimes~\cite{hobbs2021equivalence,stephani2009exact}.
Then the Coulomb gauge condition is written as
\begin{equation}
\label{eq:Coulomb_A}
 \nabla_\perp^\mu A^h_\mu = 0 
\end{equation}
with $v_\perp^\mu = \Delta^{\mu\nu} v_\nu$ and $\Delta^{\mu\nu} = g^{\mu\nu} - n^\mu n^\nu$ for a vector $v^\mu$.
In the following discussion, we focus on the curved spacetime where $g_{\mu\nu}$ satisfies Eq.~\eqref{eq:nabla_n}.
Using Eqs.~\eqref{eq:nabla_n} and~\eqref{eq:Coulomb_A}, we solve the transport equation~\eqref{eq:Weq} in a parallel manner to that in flat spacetime, up to the replacement $\partial_\mu \to D_\mu$ and the Minkowski metric $\eta_{\mu\nu}\to g_{\mu\nu}$.

For the Wigner function, the above gauge condition is represented as
\begin{eqnarray}
\label{eq:Coulomb_W1}
 &\dis \biggl( p_\perp^\lambda + \frac{\rmi \hbar}{2} D_\perp^\lambda\biggr) W_{\lambda\tau} = 0 \,,\\
\label{eq:nW}
 & n^\lambda W_{\lambda\tau}^h = 0 \,.
\end{eqnarray}
Thanks to the Hermiticity property of $W_{\mu\nu}$, we can carry out the following decomposition:
\begin{equation}
 W^h_{\mu\nu} = \calS^h_{\mu\nu} + \rmi \calA^h_{\mu\nu}\,,
 \quad
 \calS^h_{\mu\nu} = W^h_{(\mu\nu)}\,,
 \quad
 \calA^h_{\mu\nu} = -\rmi W^h_{[\mu\nu]} \,.
\end{equation}
The gauge conditions~\eqref{eq:Coulomb_W1} and~\eqref{eq:nW} thus lead to
\begin{eqnarray}
\label{eq:pDWSA1}
&\dis
 p_\perp^\lambda \calS^h_{\lambda\tau} - \frac{\hbar}{2} D^\lambda_\perp \calA^h_{\lambda\tau} = 0\,,\\
\label{eq:pDWSA2}
&\dis
 p_\perp^\lambda \calA^h_{\lambda\tau} + \frac{\hbar}{2} D^\lambda_\perp \calS^h_{\lambda\tau} = 0 \,,\\
\label{eq:nWSA}
&\dis
 n^\lambda \calS^h_{\lambda\tau} = n^\lambda \calA^h_{\lambda\tau} = 0 \,.
\end{eqnarray}
Decomposing the real and imaginary parts in Eq.~\eqref{eq:Weq}, we arrive at
\begin{eqnarray}
\label{eq:ceqSA}
 &\dis
\biggl(
 		p^2
 		- \frac{\hbar^2}{4} D^2
 	\biggr) \calS^h_{\mu\nu}
 = \biggl(
 		p^2
 		- \frac{\hbar^2}{4} D^2
 	\biggr) \calA^h_{\mu\nu}
 = 0 \,, \\
\label{eq:keqSA}
 &\dis
 p\cdot D \calS^h_{\mu\nu} = p\cdot D\calA^h_{\mu\nu} = 0 \,.
\end{eqnarray}

In general, the solutions for these equations are derived through the quantization of polarized photon fields.
Because of our gauge condition with Eq.~\eqref{eq:nabla_n}, however, it is legitimate to extract the solution in curved spacetime from that in flat spacetime.
Therefore, the general form of the Wigner function reads $ W^{h}_{\mu \nu} = \calS^{h}_{\mu\nu} + \rmi \calA^{h}_{\mu\nu}$ with~\cite{Hattori:2020gqh}
\begin{equation}
\begin{split}
\label{eq:solution}
 \calS^{h}_{\mu\nu}
 &= \pi \delta (p^2) \operatorname{sgn}(p \cdot n)
 		\left(
 			P_{\mu \nu}
 			\mp \frac{\hbar p_{\perp(\mu} S_{\nu) \alpha} D^{\alpha}}{(p \cdot n)^{2}} 
 		\right)f_{h} \,,\\
 \calA^{h}_{\mu\nu}
  &= \pi \delta (p^2) \operatorname{sgn}(p \cdot n)
		\left(
 			\mp S_{\mu \nu} 
 			- \frac{\hbar p_{\perp[\mu} D_{\perp \nu]}}{(p \cdot n)^{2}} 
 		\right) f_{h} \,,
\end{split}
\end{equation}
where we introduce $P_{\mu\nu} = -\Delta_{\mu\nu} - \hp_{\perp\mu}\hp_{\perp\nu}$ with $\hp_{\perp\mu} = p_{\perp\mu}/{\sqrt{|p_\perp\cdot p_\perp|}}$ and the spin tensor defined as
\begin{equation}
 S^{\mu\nu} = \frac{\varepsilon^{\mu\nu\rho\sigma} p_\rho n_\sigma}{p\cdot n}
\end{equation}
with the Levi-Civita tensor defined as $\varepsilon^{0123} = [-g(x)]^{-1/2}$.
Here the helicity dependence appears in two parts.
One is the distribution function $f_h$ for $h = {\rm R, L}$.
The other is the sign in front of the spin tensor $S_{\mu\nu}$: the upper and lower signs correspond to $h={\rm R, L}$, respectively.
We can check that the solutions~\eqref{eq:solution} satisfy the constraints~\eqref{eq:pDWSA1}--\eqref{eq:nWSA}.
We note that Eqs.~\eqref{eq:ceqSA} represent the on-shell condition.
Also Eq.~\eqref{eq:keqSA} corresponds to the kinetic equation for $f_{h}$ up to $O(\hbar)$, which is equivalent to the collisionless Boltzmann equation: $\delta(p^2) p^\mu(\partial_\mu - \Gamma_{\mu\nu}^\rho p_\rho\partial_\nu^p)f_{h} = 0$.

\section{Photonic currents}\label{sec:currents}
Using Eq.~\eqref{eq:solution}, we evaluate several quantities induced by photonic transport under a gravitational field.
We start from the Chern-Simons current, which is defined by
\begin{equation}\label{eq:Kh}
 \calK_h^\mu 
 = \frac{1}{2} \varepsilon^{\mu\nu\alpha\beta} 
 		\Bigl[
 			A^h_\nu \nabla_\alpha A^h_\beta
 			-(\nabla_\alpha A^h_\nu)  A^h_\beta 
 		\Bigr] \,.
\end{equation}
The Wigner transformation of Eq.~\eqref{eq:Kh} is computed as
\begin{equation}
\begin{split}
\label{eq:K_Wig}
 \calK_h^\mu (x,p) 
 &= \frac{1}{2} \varepsilon^{\mu\nu\alpha\beta}
 	\int_y \rme^{-\rmi p\cdot y/\hbar}  
 		\Bigl\langle
 			A^h_\nu (x,\tfrac{y}{2}) D_\alpha A^h_\beta(x,-\tfrac{y}{2})
 			-(D_\alpha A^h_\nu(x,\tfrac{y}{2}) )  A^h_\beta (x,-\tfrac{y}{2})
 		\Bigr\rangle \\
 &= \frac{1}{\hbar}\varepsilon^{\mu\nu\alpha\beta} p_\alpha \calA^h_{\beta\nu}(x,p) \,.
\end{split}
\end{equation}
Here we used the following relations derived from Eq.~\eqref{eq:delyA}:
\begin{equation}
\begin{split}
  A_\alpha (x,\tfrac{y}{2}) D_\mu A_\beta (x,-\tfrac{y}{2})
 &= \left(
 	- \partial_\mu^y + \frac{1}{2} D_\mu
 	\right)
 	\left[ A_\alpha (x,\tfrac{y}{2}) A_\beta (x,-\tfrac{y}{2})\right] \,, \\
  \left(D_\mu A_\alpha (x,\tfrac{y}{2}) \right) A_\beta (x,-\tfrac{y}{2})
 &= \left(
 	\partial_\mu^y + \frac{1}{2} D_\mu
 	\right)
 	\left[ A_\alpha (x,\tfrac{y}{2}) A_\beta (x,-\tfrac{y}{2})\right] \,. \\
\end{split}
\end{equation}
Performing the momentum integral of Eq.~(\ref{eq:K_Wig}) and plugging Eq.~\eqref{eq:solution} into it, 
we obtain
\begin{equation}
\begin{split}
\label{eq:K}
 K_h^\alpha(x)
 &=\int_p \calK_h^\alpha(x,p) \\
 &=\int_p \frac{\pi\delta(p^{2})}{\hbar}\operatorname{sgn}(p\cdot n)
 	(\pm 2p^{\alpha} +\hbar S^{\alpha \beta} D_\beta )f_{h} \,,
\end{split}
\end{equation}
where $\int_p = \int \frac{\rmd^4p}{(2\pi)^4} \frac{1}{\sqrt{-g}}$.

We also define the spin-3 zilch tensor as
\begin{equation}
\begin{split}
 \calZ_{\mu \nu \rho}^h 
 = \frac{1}{2} g^{\alpha\beta}
 	\left[
 		F_{\mu\alpha}^h \nabla_{\rho} \tilde{F}^h_{\nu \beta}
 		-\left(\nabla_{\rho} F^h_{\nu\alpha}\right) \tilde{F}^h_{\mu \beta}
 	\right]
\end{split}
\end{equation}
with $\tilde{F}_{\mu\nu} = \frac{1}{2} \varepsilon_{\mu\nu\rho\sigma} F^{\rho\sigma}$.
Its Wigner transformation reads
\begin{equation}
\begin{split}
\label{eq:Z3}
 \calZ_{\mu\nu\rho}^h (x,p)
 &= \frac{1}{2}
 	\int_y \rme^{-\rmi p\cdot y/\hbar} g^{\alpha\beta}
 	\left\langle
 		F^h_{\mu\alpha}(x,\tfrac{y}{2}) D_{\rho} \tilde{F}^h_{\nu \beta}(x,-\tfrac{y}{2})
 		-\left(D_{\rho} F^h_{\nu\alpha}(x,\tfrac{y}{2})\right) \tilde{F}^h_{\mu \beta}(x,-\tfrac{y}{2})
 	\right\rangle \\
 &= \frac{\rmi}{\hbar^3}p_\rho \Xi^h_{(\mu\nu)}
 	+ \frac{1}{2\hbar^2} D_\rho \Xi^h_{[\mu\nu]} \,,
\end{split}
\end{equation}
where we define 
\begin{equation}
\begin{split}
 \Xi^h_{\mu\nu}
 &=
 \varepsilon_{\mu\alpha\sigma\lambda}
 		\Biggl[
 			-\biggl(
 				p_{\nu}p^\alpha
 				+ \frac{\rmi\hbar}{2}
 					\bigl(p_{\nu} D^\alpha - p^\alpha D_{\nu}\bigr)
 			\biggr) W_h^{\sigma\lambda}
 			+ \rmi\hbar p^\alpha D^\sigma W^\lambda_{h\,\nu}
 		\Biggr] \,.
\end{split}
\end{equation}
The second term in the last line of Eq.~\eqref{eq:Z3} is a higher-order correction to the first.
This is checked from the Schouten identity and Eqs.~\eqref{eq:pDWSA1}--\eqref{eq:keqSA}.
Using Eq.~\eqref{eq:solution}, we then obtain 
\begin{equation}
\label{eq:Z3_reduced}
 \calZ^{\mu\nu\rho}_h
 = \frac{\pi\delta(p^2)}{\hbar^3}
 	\operatorname{sgn}(p\cdot n)
 	\Bigl(
 		\pm 2 p^\mu p^\nu f_h
 		+ 2 \hbar p^{(\mu} S^{\nu)\lambda} D_\lambda f_h
 	\Bigr) p^\rho \,,
\end{equation}
and the zilch current
\begin{equation}
\begin{split}
\label{eq:Z}
 Z_h^{\alpha}(x) 
 &= \int_p \Delta^{\alpha\mu} n^\nu n^\rho \calZ^h_{\mu\nu\rho}(x,p) \\
 &= \int_p \frac{\pi\delta(p^{2})}{\hbar^{3}} \operatorname{sgn}(p\cdot n)(p\cdot n)^{2} 
 		(\pm 2p_{\perp}^{\alpha} +\hbar S^{\alpha \beta} D_\beta ) f_h \,.
\end{split}
\end{equation}

In the same way, the energy-momentum tensor is evaluated as
\begin{equation}
\begin{split}
 \calT^h_{\mu\nu} (x,p)
 &= \int_y \rme^{-\rmi p\cdot y/\hbar} 
 	\biggl\langle 
 		- g^{\alpha\beta} F_{(\nu \alpha} (x,\tfrac{y}{2}) F_{\mu)\beta} (x,-\tfrac{y}{2} )
 		+\frac{1}{4} g_{\mu \nu} F^{\alpha\beta} (x,\tfrac{y}{2}) F_{\alpha\beta}(x,-\tfrac{y}{2})
 	\biggr\rangle \\
 &=  \frac{4}{\hbar^2}
	\biggl(
 		- g_{\beta(\nu} g_{\mu)\gamma} 
 		+ \frac{1}{4}  g_{\mu \nu} g_{\beta\gamma} 
 	\biggr)
 	g^{\gamma\delta}
 	\calD^{*[\alpha} \calD_{[\alpha} {W_{\delta]}}^{\beta]} \,,
\end{split}
\end{equation}
where we introduce $\calD_\mu = p_\mu + \frac{\rmi \hbar}{2} D_\mu$.
Decomposing $W_{\alpha\beta}$ into the symmetric and antisymmetric parts~\eqref{eq:solution}, we get the energy-momentum tensor as follows:
\begin{equation}
\begin{split}
\label{eq:T}
 T^h_{\mu\nu} (x)
 &= \int_p \calT^h_{\mu\nu} (x,p) \\
 & =  \int_p
 	 \frac{\pi \delta (p^2)}{\hbar^2} \operatorname{sgn}(p \cdot n)
 		(
 			2 p_\mu p_\nu
			\pm 2\hbar  p_{(\mu} S_{\nu)\alpha} D^\alpha
		)  f_{h} \,.
\end{split}
\end{equation}

%\section{Equilibrium}
Let us here briefly argue the equilibrium state.
An exact form of $f_h$ at equilibrium is found only by taking into account the collision term.
As in the fermionic cases, however, we may plausibly anticipate that the equilibrium distribution is the Bose distribution function involving the spin-vorticity coupling:
\begin{eqnarray}
\label{eq:fheq}
 &\dis f_h
 = N (g)\,,
 \quad
 g = p\cdot U 
 	\pm \frac{\hbar}{2} S^{\mu\nu} \nabla_\mu U_\nu \,,
\end{eqnarray}
augmented by the Killing condition,
\begin{eqnarray}
\label{eq:Killingeq}
\nabla_{(\lambda} U_{\alpha)} = 0,
\end{eqnarray}	
where $N(x) = (\rme^x -1)^{-1}$, $U^{\mu}=\beta u^{\mu}$ with $\beta$ the inverse of temperature $T$ and $u^{\mu}$ the fluid four velocity, and the sign $+, -$ corresponds to $h=\mathrm{R, L}$.
Under the Killing condition, the kinetic equation~\eqref{eq:keqSA} holds, and thus Eq.~\eqref{eq:fheq} is a local equilibrium distribution.%
\footnote{
Such a distribution under Eq.~\eqref{eq:Killingeq} is sometimes referred to as the ``global'' equilibrium, especially in the context of heavy-ion collision physics. In this paper, however, we call it the ``local'' equilibrium to match the terminology in the conventional thermodynamics and hydrodynamics.
}
At the same time, Eq.~\eqref{eq:Killingeq} is necessary in order to keep the frame independence of physical quantities.
For example, by making the decomposition, $\calZ_{h}^{\mu\nu\rho}=\calZ_{h(0)}^{\mu\nu\rho}+\calZ_{h\one}^{\mu\nu\rho}$ with the subscripts $(0)$ and $(1)$ corresponding to the leading order and next-to-leading order in the $\hbar$ expansion, one can show that the zilch tensor~\eqref{eq:Z3_reduced} becomes independent of the frame vector, as its $O(\hbar)$ part is written as
\begin{equation}
\begin{split}
\label{eq:calZ1}
 \calZ_{h\one}^{\mu\nu\rho}
 & = \frac{\pi\delta(p^2)}{\hbar^2}
 	\operatorname{sgn}(p\cdot n) N'(p\cdot U) 
 	\left[
 		 p^\mu p^\nu
 		 S^{\alpha\beta} \nabla_\alpha U_\beta
 		+ 2 p^{(\mu} S^{\nu)\lambda} D_\lambda (p\cdot U)
 	\right] p^\rho \\
 & = \frac{\pi\delta(p^2)}{\hbar^2}
 	\operatorname{sgn}(p\cdot n) N'(p\cdot U) 
 		\nabla_\lambda U_\alpha \,
 		\varepsilon^{\lambda\alpha\sigma(\mu}p^{\nu)} p_\sigma p^\rho \,.
\end{split}
\end{equation} 
Here we used Eq.~\eqref{eq:Killingeq} and
\begin{equation}
 S^{\mu[\alpha}p^{\beta]}
 = -\frac{1}{2} S^{\alpha\beta}p^\mu
 	- \frac{1}{2} \varepsilon^{\alpha\beta\mu\nu}
 	\biggl(
 		p_\nu
 		- \frac{p^2 n_\nu}{p\cdot n}
 	\biggr) \,,
\end{equation}
which follows from the Schouten identity.
A similar computation also leads to the frame independence of the energy-momentum tensor~\eqref{eq:T}.

\subsection{Chiral/zilch vortical effect}
One of the most instructive applications is the photon transport phenomena under rotation.
We here consider the slowly rotating coordinate, which is described by
\begin{equation}
\label{eq:g_rot}
g_{\mu\nu} = \eta_{\mu\nu} + h_{\mu\nu}\,,
\quad
|h_{\mu\nu}|\ll 1\,,
\quad
h_{0i} = h_{0i}(\bx) \,,
\quad
h_{00} = h_{ij} = 0 \,.
\end{equation}
In the following we ignore $O(h^2)$.
For this metric tensor, the frame vector~\eqref{eq:frame} reads
\begin{equation}
 n_\mu = (1,\bzero)\,.
\end{equation}
The analysis in the previous section is applicable because the rest frame condition~\eqref{eq:nabla_n} holds.

We adopt the equilibrium state~\eqref{eq:fheq} with 
\begin{equation}
u^{\mu}=\delta^{\mu}_0,
\end{equation}
such that $u_{i}=h_{0i}({\bx})$ and $\beta$ being a constant.
This satisfies the Killing condition~\eqref{eq:Killingeq}.
Physically, such a fluid velocity represents a fluid corotating with the coordinate described by Eq.~\eqref{eq:g_rot}.
As the zilch tensor~\eqref{eq:calZ1}, we evaluate the Chern-Simons current~\eqref{eq:K}, zilch current~\eqref{eq:Z} and energy momentum tensor~\eqref{eq:T} for $h = {\rm R, L}$ as
\begin{equation}
\begin{split}
\label{eq:PVE}
K^\alpha_{h(1)}(x)
&= -\frac{\beta\omega^\alpha}{3\pi^2}\int_0^\infty \rmd|\bp| |\bp|^2 N'(\beta|\bp|) 
= \frac{1}{9}T^2\omega^\alpha \,,
\end{split}
\end{equation}
\begin{equation}
\begin{split}
\label{eq:ZVE}
Z_{h(1)}^{\alpha}(x)
&= -\frac{\beta\omega^\alpha}{3\pi^2 \hbar^{2}}\int_0^\infty \rmd|\bp| |\bp|^4 N'(\beta|\bp|) 
= \frac{4 \pi^2}{45 \hbar^{2}} T^4 \omega^\alpha \,,
\end{split}
\end{equation}
\begin{equation}
\begin{split}
\label{eq:EVE}
T_{h(1)}^{\mu\nu}(x)
&= \mp \frac{2\beta u^{(\mu}\omega^{\nu)}}{3\pi^2 \hbar}\int_0^\infty \rmd|\bp| |\bp|^3 N'(\beta|\bp|) 
= \pm \frac{4\zeta(3)}{\pi^2\hbar} T^3 u^{(\mu}\omega^{\nu)} \,,
\end{split}
\end{equation}
up to $O(h_{\mu\nu})$.
Here we again use the subscript ${(1)}$ to represent the $O(\hbar)$ correction for different quantities and introduce the four-vorticity
\begin{equation}
\label{eq:omega}
\omega^\mu = \frac{1}{2} \varepsilon^{\mu\nu\rho\sigma} u_\nu\nabla_\rho u_\sigma \,.
\end{equation}
Therefore, the photonic quantum kinetic theory in the rotating coordinate correctly reproduces the photonic chiral vortical effect~\cite{Avkhadiev:2017fxj,Yamamoto:2017uul,PhysRevA.96.043830,Huang:2018aly,Prokhorov:2020okl} and the zilch vortical effect~\cite{Chernodub:2018era,Copetti:2018mxw}.
This fact justifies the validity of the present framework.
We note that the above vortical currents consist of two contributions.
One is the first terms in Eqs.~\eqref{eq:K}, \eqref{eq:Z} and \eqref{eq:T}, originating from the energy correction due to the spin-vorticity coupling $\Delta\epsilon_h^\text{rot} = \pm \frac{\hbar}{2} S^{\mu\nu} \nabla_\mu U_\nu \sim \pm \hbar \hat{\bp}\cdot\bomega$.
Another is the second terms corresponding to the magnetization current $\hbar S^{\mu\nu} D_\nu f_h$.

\subsection{Spin Hall effect for helicity/energy currents}
Let us now derive the nontrivial transport phenomena induced by background gravitational fields, which is the main part of this paper.
We consider the coordinate system described by
\begin{equation}
\label{eq:g_hall}
g_{\mu\nu} = \eta_{\mu\nu} + h_{\mu\nu} \,,
\quad
 |h_{\mu\nu}|\ll 1\,,
\quad
 h_{ij} = 4\delta_{ij} \phi(\bx)\,,
 \quad
 h_{0\mu} = 0 \,,
\end{equation}
where $\eta_{\mu\nu}=\text{diag}(1,-1,-1,-1)$,
with which the metric is
\begin{equation}
\rmd s^2 = \rmd t^2 - (1-4\phi)\rmd \bx^2 \,.
\end{equation}
This geometry can be treated as a medium effect on light~\cite{carroll2019spacetime}.
Indeed the refractive index is deviated from unity as%
\footnote{Although the parametrization of the metric perturbation differs from that in Ref.~\cite{Yamamoto:2017gla}, both of them lead to the same refractive index.}
\begin{equation}
\label{eq:ref_ind}
 n(\bx)= \biggl(\frac{|\rmd\bx|}{\rmd t}\biggr)^{-1} \simeq 1-2\phi(\bx) \,,
\end{equation}
where the null geodesic equation $\rmd s^2 = 0$ is imposed.
The frame vector~\eqref{eq:frame} is
\begin{equation}
 n_\mu = (1,\bzero)\,,
\end{equation}
for which Eq.~\eqref{eq:nabla_n} is satisfied.

We again employ the equilibrium state~\eqref{eq:fheq}.
The above metric tensor admits
\begin{equation}
 U^\mu = \beta (1,\bu) \,,
 \quad
 |\bu| \ll 1 \,,
\end{equation}
as a Killing vector, if the following conditions are fulfilled:
\begin{equation}
 \partial_{(i} u_{j)} = 0 \,,\quad
 \bu\cdot\bnabla \phi = 0 \,,
\end{equation}
where $\bu\cdot\bnabla := u^{i}\partial_i$.
The former condition implies that the fluid is shear-free.
The latter means that $\phi$ is static at this fluid frame: $D\phi/Dt:=\partial_t \phi + \bu\cdot\bnabla \phi = 0$.
Hereafter we impose the above conditions on $\bu$.
Then, the Chern-Simons current, zilch current, and energy-momentum tensor are given by totally the same forms as Eqs.~\eqref{eq:PVE}--\eqref{eq:EVE}, respectively.
In the present coordinate, however, the fluid vorticity field~\eqref{eq:omega} involves two types of contributions:
\begin{equation}
\label{eq:omega_phi}
 \bomega = \tilde{\bomega} - 2\bnabla\phi\times\bu 
\end{equation}
with $\tilde{\bomega} := \frac{1}{2}(1+2\phi)\bnabla\times\bu$.
Here we used $\varepsilon^{0123} = (-g)^{-1/2} = 1+6\phi$ for the metric~\eqref{eq:g_hall}.
For the energy dispersion of semiclassical particles, the first term in Eq.~\eqref{eq:omega_phi} gives the usual spin-vorticity coupling $\Delta \epsilon_h^\text{vort} = \pm\hbar\hat{\bp}\cdot \tilde{\bomega}$, and thus leads to the vortical currents through Eqs.~\eqref{eq:PVE}--\eqref{eq:EVE}.
The second yields an extra energy correction due to spin and inhomogeneous refraction:
\begin{equation}
\label{eq:epsilon_ref}
 \Delta \epsilon_h^\text{ref}
 = \mp\,2\hbar\hat{\bp}\cdot(\bnabla\phi\times\bu) = \pm \hbar\hat{\bp}\cdot (\bnabla n \times\bu) \,.
\end{equation}
This is responsible for photonic Hall currents, which are obtained via the replacement $\omega^i\to -2(\bnabla\phi\times\bu)^i$ in Eqs.~\eqref{eq:PVE}--\eqref{eq:EVE} as
\begin{eqnarray}
\label{eq:bK_Hall}
\dis
 \bK_{h\,\text{Hall}}
 &=& -\frac{2}{9}T^2\,\bnabla\phi\times \bu \,, \\
\label{eq:bZ_Hall}
\dis
 \bZ_{h\,\text{Hall}}
 &=& -\frac{8\pi^2}{45\hbar^2}T^4 \,\bnabla\phi\times \bu \,,\\
\label{eq:bJ_Hall}
\dis
 \bJ_{h\,\text{Hall}}
 &=& \mp\frac{8\zeta(3)}{\pi^2\hbar}T^3 \,\bnabla\phi\times \bu \,,
\end{eqnarray}
with $J_h^\alpha := T^{\alpha 0}_h$.
The currents~\eqref{eq:bK_Hall} and~\eqref{eq:bZ_Hall} represent the spin Hall effect for the refractive index~\eqref{eq:ref_ind}.
On the other hand, Eq.~\eqref{eq:bJ_Hall} represents the spin Hall energy current, which is first derived here to our knowledge. Unlike $\bK_{h\,\text{Hall}}$ and $\bZ_{h\,\text{Hall}}$, this energy current has the opposite sign for $h={\rm R,L}$.
The spin Hall energy current is hence realized only in polarized photonic systems.

It is obvious from the above derivation that the algebraic structures of the Hall currents~\eqref{eq:bK_Hall}--\eqref{eq:bJ_Hall} are totally the same as the corresponding vortical currents.
The Hall currents consist of two contributions, as do the vortical currents: one from the spin-refraction coupling $\Delta\epsilon_h^\text{ref} = \mp\,2\hbar\hat{\bp}\cdot(\bnabla\phi\times\bu)$ and the other from the magnetization current $\hbar S^{\alpha\beta} D_\beta f_h$.
Besides, the transport coefficients for the spin Hall effect should match those of the vortical effect.

As a consistency check, let us now recall the semiclassical one-particle kinetics with the Berry curvature.
The photonic current is obtained as the integral
\begin{equation}
\label{eq:bK_xdot}
 \bK_h = \pm\int_\bp \dot{\bx} N(\beta\epsilon_h) \,,
\end{equation}
where $\dot{\bx}$ and $\epsilon_h$ are the velocity and energy dispersion of the polarized photons.
In a medium with the reflective index~\eqref{eq:ref_ind}, the semiclassical equation of motion is given by~\cite{Gosselin:2006wp,Yamamoto:2017gla}
\begin{equation}
 \dot{\bx} = \frac{\hat{\bp}}{n} \pm 2\hbar\bnabla\phi\times \frac{\hat{\bp}}{|\bp|} \,,
\end{equation}
where the second term emerges due to the Berry curvature.
From the classical on-shell condition $0=p^2 = g^{\mu\nu} p_\mu p_\nu$ with the correction in Eq.~\eqref{eq:epsilon_ref}, the energy dispersion is identified as
\begin{equation}
\begin{split}
 \epsilon_h
 &= p_0 + p_i u^i + \Delta\epsilon_h^\text{ref} \\  
 &= (1+2\phi)|\bp|-\bu\cdot\bp \mp \,2\hbar\hat{\bp}\cdot(\bnabla\phi\times\bu)
\end{split}
\end{equation}
with $\bp = (-p_1,-p_2,-p_3)$.
Plugging these two pieces into Eq.~\eqref{eq:bK_xdot}, we evaluate the $O(\hbar)$ part as
\begin{equation}
\begin{split}
 \bK_{h\,\one}
 &\simeq 
 	\int_\bp 
 	\biggl[
 		-2\hat{\bp}\Bigl(\hat{\bp}\cdot (\bnabla\phi\times\bu)\Bigr)
 		- 2\bnabla\phi\times \frac{\hat{\bp}}{|\bp|} (\bu\cdot\bp)
 	\biggr]
 		N'(\beta|\bp|) \\
 &=-\frac{2}{9}T^2\,\bnabla\phi\times \bu \,,
\end{split}
\end{equation}
where we keep $O(\bu)$ and $O(\phi)$.
One thus arrives at the spin Hall current in Eq.~\eqref{eq:bK_Hall} through a different derivation.

\section{Conclusion and Discussion}\label{sec:discussion}
In this paper, we derived the photonic quantum kinetic theory in curved spacetime from quantum field theory, as an extension of Ref.~\cite{Hattori:2020gqh}.
This framework reproduces the photonic/zilch vortical effect in a rotating coordinate.
We showed that the photonic Wigner function yields the photonic spin Hall effect for the helicity/energy current due to the Berry curvature in an external gravitational field. In particular, the latter transport phenomenon was computed for the first time in this paper.
One of crucial findings is that the photonic spin Hall effect and vortical effect have totally the same origin.
It indicates that the photonic Hall effect might also be categorized in anomalous transport phenomena dictated by quantum anomalies, if so are the vortical ones~\cite{Dolgov:1988qx,Agullo:2016lkj}.
Such a matching relation is found from the general covariant form of the currents~\eqref{eq:PVE}--\eqref{eq:EVE}.
This is an advantage of the present work compared to the noncovariant theory employed in usual analysis for the photonic Hall effect~\cite{PhysRevLett.93.083901,Berard:2004xn,Gosselin:2006wp,BLIOKH2004181,Duval:2005ky,Yamamoto:2017gla}.
It should be emphasized that although we adopted a specific frame vector defining the spin polarization of photons, our results do not depend on the choice of the frame vector, as in the fermionic case~\cite{Chen:2014cla,Chen:2015gta,Hidaka:2017auj};
while the position of massless photons is generally frame dependent~\cite{Harte:2022dpo}, the physical currents such as the zilch currents and spin Hall energy current are not.

The essential ingredient of this work is the frame-vector condition~\eqref{eq:nabla_n}.
Several coordinates, such as the rotating one with Eq.~\eqref{eq:g_rot} and the spatially deformed one with Eq.~\eqref{eq:g_hall} satisfy this condition.
This is not the case for some curved spacetime.
For instance, when we consider a spatially dependent $g^{00}(\bx)$, the Wigner functions~\eqref{eq:solution} receive extra contributions from $\nabla_\mu n_\nu \neq 0$, and so do the Chern-Simons and zilch currents.
This class of coordinate is relevant to realistic physical environments, e.g., the Schwarzschild black hole in astrophysics and thermal gradient in condensed matter systems.
Nonetheless, this should be a technical difficulty that could be potentially resolved by choosing a proper frame vector.
The photonic quantum kinetic theory in more general curved spacetime will be investigated in the future work.

Furthermore, from the transport equation~\eqref{eq:Weq-full}, we can study the photonic quantum transport theory with higher-order quantum corrections.
In the Wigner functions up to $O(\hbar^2)$, there emerges the Riemann curvature effect.
Coupled with the photon spin, the curvature could bring not only a contribution to gravitational lensing~\cite{Bartelmann:1999yn,Bartelmann:2010fz} but also novel nondissipative transport phenomena in the context of condensed matter systems and heavy-ion collisions~\cite{Hayata:2020sqz}.

\begin{acknowledgments}
K.~M. was supported by Special Postdoctoral Researcher (SPDR) Program of RIKEN.
N.~Y. was supported by the Keio Institute of Pure and Applied Sciences (KiPAS) project at Keio University and JSPS KAKENHI Grant No.~19K03852.
D.-L. Y. was supported by the Ministry of Science and Technology, Taiwan under Grant No. MOST 110-2112-M-001-070-MY3.
\end{acknowledgments}

\appendix

\section{Transport equation}\label{app:transport}
In this Appendix, we derive the transport equation for the photonic Wigner function $W_{\mu\nu}(x,p)$ in general coordinate.
The Riemann curvature is defined as ${R^\rho}_{\sigma \mu \nu}=2(\partial_{[\nu} \Gamma_{\mu] \sigma}^{\rho}+\Gamma_{\lambda[\nu}^{\rho} \Gamma_{\mu] \sigma}^{\lambda})$ and the Ricci tensor is $R_{\mu\nu} = {R^\rho}_{\mu\rho\nu}$.
We focus on the free photon field operators that obey the Maxwell's equation under gravity:
\begin{equation}
\begin{split}
 \label{eq:Maxwell}
 0&=\nabla^\mu F_{\mu\nu} 
 = \nabla^2 A_\nu - \nabla_\nu \nabla\cdot A
 	+ R_{\nu\lambda} A^\lambda \,,
\end{split}
\end{equation}
where we use $[\nabla_\mu,\nabla_\nu]A_\rho = - R_{\mu\nu\rho\lambda} A^\lambda$.
It is here useful to prepare the operator identity
\begin{equation}
\label{eq:Oexp}
O \rme^{y\cdot D} = \rme^{y\cdot D} O + (1-\rme^{\calC(y\cdot D)})O \rme^{y\cdot D} \,,
\quad
\calC(Y) Z = [Y,Z] \,,
\end{equation}
which follows from $ \rme^{Y} X \rme^{-Y} = \rme^{\calC(Y)}X$.
For the derivative operators, Eq.~\eqref{eq:Oexp} implies
\begin{eqnarray}
 \label{eq:delyA}
 &\dis
 \partial^y_\mu A_\lambda(x,y) 
 = (D_\mu  + 2\calG_\mu ) A_\lambda(x,y) \,, \\
 \label{eq:DA}
 &\dis
 D_\mu \rme^{y\cdot D} 
 = \rme^{y\cdot D} \nabla_\mu - \calH_\mu \rme^{y\cdot D}\,, \\
 \label{eq:DDA}
 & \dis
 D_{(\mu} D_{\nu)} A_\lambda(x,y) 
  = \Bigl[
  		(\rme^{y\cdot D}\nabla_{(\mu} - \calH_{(\mu}\rme^{y\cdot D} )
  			 \nabla_{\nu)} 
  		-  D_{(\mu} \calH_{\nu)} \rme^{y\cdot D}
  	  \Bigr] A_\lambda (x) \,,
\end{eqnarray}
where we introduce the shorthand notations for the following operators:
\begin{eqnarray}
 &\dis
 \calG_\mu 
  = -\frac{\rmi y^\nu}{2\hbar}\sum_{n=0}^\infty\frac{\bigl[\calC(y\cdot D)\bigr]^n}{(n+2)!} 
  \, G_{\mu\nu} \,, \\
 &\dis
  \calH_\mu 
  = -\frac{\rmi y^\nu}{\hbar}\sum_{n=0}^\infty\frac{\bigl[\calC(y\cdot D)\bigr]^n}{(n+1)!} 
  \, G_{\mu\nu} \,, \\
  &\dis
  G_{\mu\nu} 
  =-\rmi\hbar [D_\mu,D_\nu] \,.
\end{eqnarray}

Let us calculate the two point correlator
\begin{equation}
\label{eq:ADDA}
 G_{\mu\nu\lambda\tau}
 = \int_y \rme^{-\rmi p\cdot y/\hbar} \biggl\langle A_\tau^+
  \biggl(\frac{1}{2} D_{(\mu} - \partial_{(\mu}^y\biggr)
  \biggl(\frac{1}{2} D_{\nu)} - \partial_{\nu)}^y\biggr)
  A_\lambda^- \biggr\rangle \,,
  \quad
  A^\pm = A(x,\pm y/2) \,.
\end{equation}
On one hand, by using Eq.~\eqref{eq:delyA}, we write
\begin{equation}
\begin{split}
\label{eq:G1}
  G_{\mu\nu\lambda\tau}
 & = \biggl(\frac{1}{2} D_{(\mu} - \frac{\rmi p_{(\mu}}{\hbar} \biggr)
  \biggl(\frac{1}{2} D_{\nu)} -\frac{\rmi p_{\nu)}}{\hbar} \biggr) W_{\lambda\tau}
  + 2\biggl(\frac{1}{2} D_{(\mu} - \frac{\rmi p_{(\mu}}{\hbar} \biggr)
  	\int_y \rme^{-\rmi p\cdot y/\hbar} \biggl\langle 
  	 \calG_{\nu)} A_\tau^+  
  	 A_\lambda^- \biggr\rangle \\
 & \quad
  + \int_y \rme^{-\rmi p\cdot y/\hbar} \biggl\langle 
  	 -\biggl[
  	 \biggl(
  	 	\frac{1}{2} D_{(\mu} - \partial_{(\mu}^y 
  	 \biggr) \calG_{\nu)} 
  	  A_\tau^+ 
  	 \biggr]
  	 A_\lambda^- \biggr\rangle \,.
\end{split}
\end{equation}
On the other hand, Eqs.~\eqref{eq:delyA} and~\eqref{eq:DDA} lead to
\begin{equation}
\begin{split}
\label{eq:G2}
  G_{\mu\nu\lambda\tau}
 & = \int_y \rme^{-\rmi p\cdot y/\hbar} \biggl\langle 
  	 A_\tau^+ 
  	 \biggl[
  	 	 \biggl(
  	 		\frac{3}{2} D_{(\mu} - \partial_{(\mu}^y 
  	 	  \biggr) 
  	 	  \calG_{\nu)} 
  	 	- \Bigl(
  			 D_{(\mu} \calH_{\nu)} 
  			 + \calH_{(\mu} D_{\nu)} 
  			 + \calH_{(\mu} \calH_{\nu)} 
 	 	  \Bigr)
  	 \biggr]
  	 A_\lambda^- \biggr\rangle \\
  &\quad
  	 + \int_y \rme^{-\rmi p\cdot y/\hbar} \biggl\langle 
  	 A_\tau^+ \rme^{-y\cdot D/2} \nabla_{(\mu}\nabla_{\nu)} A_\lambda (x) \biggr\rangle \,.
\end{split}
\end{equation}
Multiplying Eqs.~\eqref{eq:G1} and~\eqref{eq:G2} by $g^{\mu[\nu}g^{\lambda]\rho}$ and using the Maxwell's equation~\eqref{eq:Maxwell}, we derive the following transport equation:
\begin{equation}
\begin{split}
 \label{eq:Weq-full}
  & g^{\mu[\nu}g^{\lambda]\rho} 
   		\biggl(p_{(\mu} + \frac{\rmi\hbar}{2} D_{(\mu}\biggr)
  		\biggl(p_{\nu)} + \frac{\rmi\hbar}{2} D_{\nu)}\biggr)
  		W_{\lambda\tau} 
  = \hbar^2 g^{\mu[\nu}g^{\lambda]\rho} 
  	\Bigl(
  		X_{\mu\nu\lambda\tau} 
  		+ Y_{\mu\nu\lambda\tau}
  	\Bigr)
  	 +\hbar^2 {Z^\rho}_\tau \,,
\end{split}
\end{equation}
where
\begin{equation}
\begin{split}
 X_{\mu\nu\lambda\tau}
 & = \int_y \rme^{-\rmi p\cdot y/\hbar}
  	 \biggl[
  	 (\partial^y_{(\mu} \calG_{\nu)})^-
  	 + (\partial^y_{(\mu} \calG_{\nu)})^+
  	 - \frac{3}{2} (D_{(\mu}\calG_{\nu)})^-
  	 - \frac{1}{2} (D_{(\mu}\calG_{\nu)})^+ \\
 & \qquad
  	 + (D_{(\mu} \calH_{\nu)} )^-
  	 + (\calH_{(\mu} D_{\nu)} )^-
  	 + (\calH_{(\mu} \calH_{\nu)} )^-
  	  \biggr]
  	   \langle  A_\tau^+ A_\lambda^- \rangle \,, \\
 Y_{\mu\nu\lambda\tau}
 &= -\frac{2\rmi}{\hbar}\biggl( p_{(\mu} + \frac{\rmi\hbar}{2} D_{(\mu}\biggr)
  	\int_y \rme^{-\rmi p\cdot y/\hbar}\, 
  	 \calG_{\nu)}^+ \langle  A_\tau^+ A_\lambda^- \rangle \,, \\
 {Z^\rho}_\tau
 &= \frac{1}{2} (\rme^{-\calC(\rmi\hbar\partial_p\cdot D/2)} R^{\rho\lambda}) W_{\lambda\tau} \,,
\end{split}
\end{equation}
with $O^{+} \langle  A_\tau^+ A_\lambda^- \rangle = \langle  (O A_\tau^+) A_\lambda^- \rangle $ and $O^{-} \langle  A_\tau^+ A_\lambda^- \rangle = \langle   A_\tau^+ (OA_\lambda^-) \rangle $.
For the flat spacetime, the curvature contributions from the right-hand side of Eq.~\eqref{eq:Weq-full} disappear, and we obtain Eq.~\eqref{eq:Weq}.

%--- Bibliography ---%
\bibliography{photon}
%--- Bibliography ---%

\end{document}